\documentclass[10pts,sigconf,letterpaper, nonacm]{acmart}

\usepackage{graphicx}
\usepackage{subcaption}

\newcommand{\figref}[1]{Fig.~\ref{fig:#1}}

\newcommand{\tabref}[1]{Table~\ref{tab:#1}}
\newcommand{\secref}[1]{Section~\ref{sec:#1}}

\renewcommand{\eqref}[1]{(\ref{eq:#1})}
\newcommand{\eg}{{\it e.g.}}
\newcommand{\ie}{{\it i.e.}}

\newcommand{\sysname}{LoLa}
\newcommand{\emuname}{CellNem}

\title{\sysname: Low-Latency Realtime Video Conferencing over Multiple Cellular Carriers}

\author{Sara Ayoubi}
\authornote{This research was conducted during their postdoctoral fellowship at Inria Paris Research Center in 2018.}
\email{sara.ayoubi@nokia-bell-labs.com}
\affiliation{%
  \institution{Nokia Bell Labs}
  \city{Paris-Saclay}
  \country{France}
}

\author{Giulio Grassi}
\authornotemark[1]
\affiliation{%
  \institution{Inria}
  \city{Paris}
  \country{France}
}

\author{Giovanni Pau}
\affiliation{%
  \institution{University of Bologna}
  \city{Bologna}
  \country{Italy}
}

\author{Kyle Jamieson}
\affiliation{%
  \institution{Princeton University}
  \city{Princeton}
  \country{USA}
}

\author{Renata Teixeira}
\email{renata.teixeira@inria.fr}
\affiliation{%
  \institution{Inria}
  \city{Paris}
  \country{France}
}

\begin{document}

\setcopyright{none}
\settopmatter{printacmref=false} 
\renewcommand\footnotetextcopyrightpermission[1]{} 
\pagestyle{plain} 

\begin{abstract}
\sysname{} is a novel multi-path system for video conferencing
applications over cellular networks. It provides
significant gains over single link solutions when the link quality over
different cellular networks fluctuate dramatically and
independently over time, or when aggregating the throughput across different
cellular links improves the perceived video quality.
\sysname{} achieves this by continuously estimating the quality of available
cellular links
to decide how to strip video packets across them without inducing delays or
packet drops. It is also tightly coupled with state-of-the-art video codec to
dynamically adapt video frame size to respond quickly to changing network
conditions.\\
Using multiple traces collected over 4 different cellular operators in a large
metropolitan city, we demonstrate that \sysname{} provides significant gains in
terms of throughput and delays compared to state-of-the-art real-time video
conferencing solution.
\end{abstract}

\maketitle

\section{Introduction}

Despite the growing prevalence of cellular networks for internet access,
users may still encounter poor network conditions caused by fluctuations in
link rates and sudden outages that can last for several seconds~\cite{yap2012making, winstein2013stochastic}.
These poor network conditions are particularly detrimental for
latency-sensitive applications such as video conferencing programs,
which require network delays of less than 100 milliseconds to maintain
the naturalness of two-way communication~\cite{recommendation1996114,winstein2013stochastic}.
\begin{figure}
    \centering
    \includegraphics[width=1\linewidth]{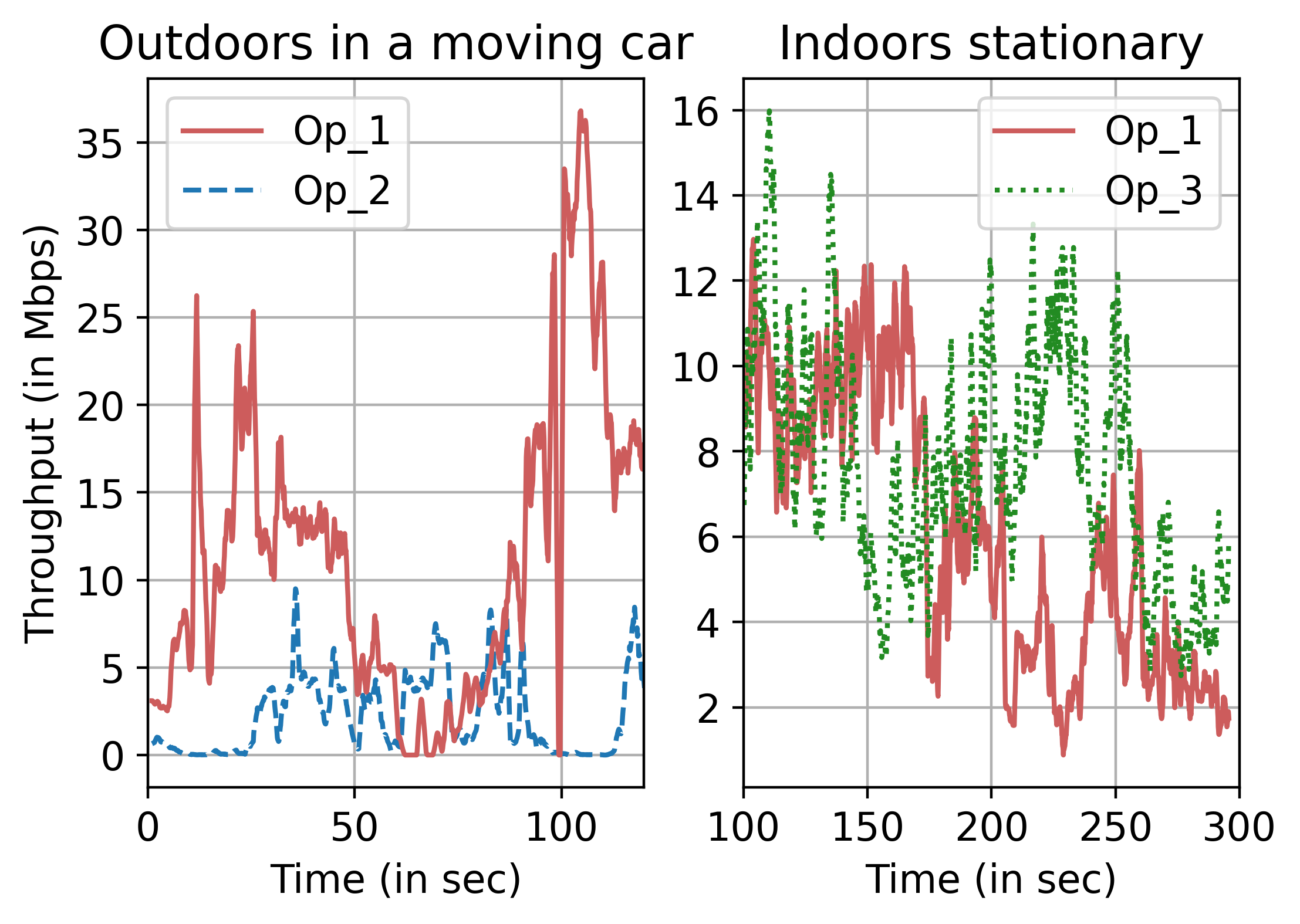}
    \caption{Uplink throughput measured on two different cellular operators
             in a large metropolitan city in indoor and outdoor settings.}
    \label{fig:fluctuations_in_cellular}
\end{figure}

Previous efforts~\cite{winstein2013stochastic,fouladi2018salsify} have
attempted to address this problem by adapting the
application's sending rate to closely match the available link rate.
However, these approaches fail to prevent lags during link outages and can
result in low-quality or frequent video-quality-level switches in the face of
rapid link rate changes.

Figure~\ref{fig:fluctuations_in_cellular} illustrates the uplink throughput
measured for two different pairs of commercial cellular operators in mobile
and stationary conditions. Our collected data confirms previous observations
of frequent rapid link rate changes and outages and demonstrates that there
is little correlation between these fluctuations across different operators.
These observations motivated us to explore the potential benefits of
utilizing multiple cellular interfaces simultaneously to offer
higher-quality video conferencing sessions. By aggregating throughput
across multiple cellular interfaces, it may be possible to stabilize or
even improve the visual quality of a video conferencing session, and
by moving packets away from a congested link in a timely fashion, we can
improve the end-to-end delay and reduce or prevent video lags.

With this in mind, we designed \sysname{}, the first multipath system for 
real-time interactive video conferencing applications. \sysname{}~is designed
for multipath transmission over multiple cellular carriers and exploits the
availability of multiple network interfaces to dynamically strip packets
across the available paths, providing the highest interactive video quality
while bounding the end-to-end network delays. We conducted a trace-driven
experimental evaluation of LoLa using data collected from four different
commercial vendors in a large metropolitan city in both indoor and outdoor
environments. Our results confirm \sysname{}'s
ability to provide higher video quality and lower latency compared to the
state-of-the-art single-trace video conferencing solution. The main 
contributions of \sysname{}~are:
\begin{itemize}
\item The first end-to-end multi-path system for real-time video
conferencing applications.
\item An improved trace-driven emulator to enable high-fidelity record and
replay of cellular network conditions.
\item A dataset of mobile and stationary network traces captured across
6 different pairs of operators in a large metropolitan city.
\end{itemize}

The rest of the paper is organized as follows: we discuss related work in 
\secref{related_work}. In \secref{system_design} we 
present the design of \sysname{}. In \secref{cellular_emulator} we detail the
evaluation tested, present our numerical results in \secref{evaluation},
and conclude the paper in \secref{conclusion}.

\section{Related Work}
\label{sec:related_work}

Several multi-path systems and protocols have been made to date,
with the most prominent being Stream Control Transmission Protocol
(SCTP)~\cite{stewart2007stream} and Multipath-TCP
(MP-TCP)~\cite{raiciu2012hard}. Both protocols support the use of multiple
interfaces to reach an endpoint, but they differ in their approach.
SCTP only uses the second path as a backup in case the primary path fails.
Extension proposals have been made~\cite{abd2004ls,iyengar2006concurrent}
to enable concurrent multipath transfer with SCTP, but it remains largely
unused in practice.

MP-TCP~\cite{raiciu2012hard} is the first standardized and widely adopted
protocol that splits application traffic across multiple
paths and uses a coupled multi-path congestion control algorithm to avoid
congested paths and achieve fairness with competing flows. However,
MP-TCP was designed to support reliable applications with loose latency
requirements. A study~\cite{nikravesh2016depth} demonstrates that when it comes
to latency sensitive applications (voice-over-IP), sessions running over MP-TCP
achieve worse perceived quality compared to single-path TCP. 
The main reason behind this performance degradation is variation in
latency across the multiple available paths which lead to high level
of jitters and out-of-order packet deliveries.

Several extensions have been proposed to optimize MP-TCP for specific use-cases.
MP-DASH~\cite{han2016mp} is an MP-TCP scheduler that
aims to minimize the use of cellular data for on-demand video streaming
applications. RAVEN~\cite{lee2018raven} is another extension that aims to
improve the performance of latency sensitive applications in connected cars.
However, MP-DASH is optimized for video streaming, which allows it to
prebuffer video frames to cushion against link variations, while
RAVEN is tailored for low-throughput applications that require reliable
transmission. \sysname{}~ is designed
to support interactive video which does not require the same level of reliability
and is more bandwidth-demanding. Further, the real-time nature of interactive
video precludes the use of video buffers.

MP-QUIC~\cite{de2017multipath} is a multipath implementation of QUIC,
a protocol designed to address slow connection setup and security limitations
in TCP. A study~\cite{zheng2021xlink}
measured the performance of vanilla MP-QUIC in a production short-video service
over wireless network traces collected in mobile environments.
Their results show that MP-QUIC achieves higher rebuffering rate and
worse 99th percentile completion time compared to single path QUIC.
This led to the development of XLINK, an MP-QUIC extension,
which focuses on improving performance for short-form video streaming in
mobile settings. However, while XLINK focuses on quality metrics such as
video request completion time and first-video-frame latency,
interactive video support requires both maximizing perceived video quality
and limiting end-to-end network delays.

EMS~\cite{chow2009ems} is an adaptive load splitting and forward error
correction schemes for multi-path real-time live streaming. However, EMS
only focuses on per-packet deadlines and is not aware of video frame deadlines
and in-order video frame deliveries. Moreover, EMS is decoupled from the codec
and assumes constant bitrate streaming. DAMS~\cite{zuo2022deadline} is a
multi-path scheduler for MP-QUIC designed
to support interactive video applications such as video conferencing and
live streaming where video frames have heterogeneous attributes (size,
priority, deadlines). Compared to DAMS and EMS, \sysname{}~is an end to
end system for interactive video over multiple
cellular carriers. In this respect, \sysname{}~is more suited to infer and
adapt to cellular network conditions as it is tightly coupled with the
video codec, and is tailored to improve support over cellular networks where
delays are mostly self-inflicted~\cite{winstein2013stochastic}.

Other research~\cite{singh2011multipath, pakulova2017adaptive} has looked at
extending the Real-time Communication Protocol (RTP) to multipath RTP
(MP-RTP). However, MP-RTP was designed for live video streaming applications,
which can tolerate delays of half a second, compared to the stricter
requirements for interactive video. Additionally, existing MP-RTP solutions
rely on the Real-Time Control Protocol (RTCP) to assess network conditions,
which generates feedback reports every second and can be slow to adapt to
rapidly changing link rates. Furthermore, the existing MP-RTP
system~\cite{pakulova2017adaptive} utilizes Scalable Video Coding (SVC) to
switch the video bitrate to available network capacities, which is a
coarse-grained approach. Our proposed solution integrates with
state-of-the-art video codec~\cite{fouladi2018salsify} that allows
frame-by-frame optimization, and a quick feedback loop, thereby achieving 
fine-grained adaptation to the available network capacity.

\section{System Design and Implementation}
\label{sec:system_design}

\subsection{System Overview}

\begin{figure*}[ht!] \begin{center}
\includegraphics[width=\linewidth]{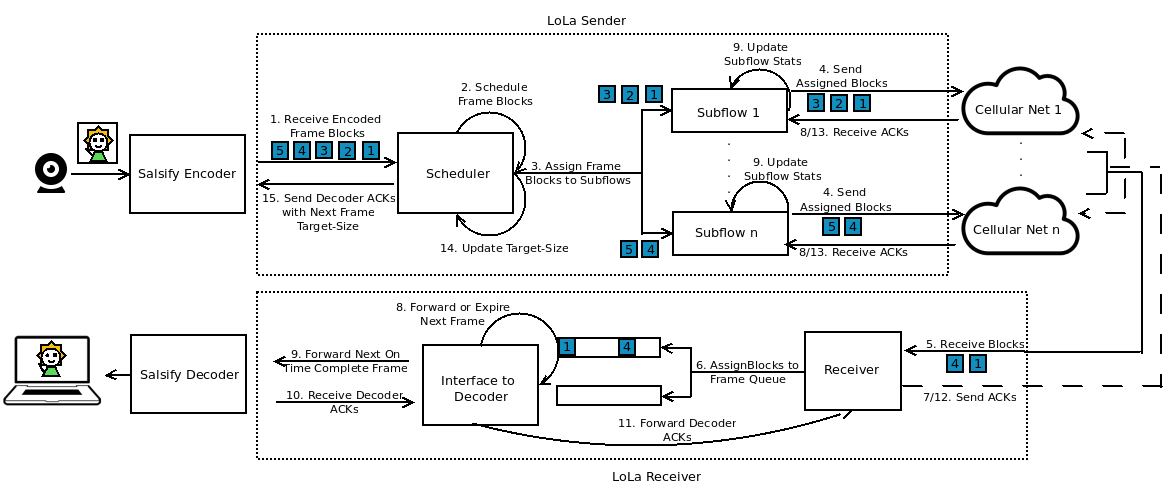}
\caption{System Architecture of \sysname{}} \label{fig:system_overview} \end{center}
\end{figure*}

An overview of \sysname{}'s architecture is provided in \figref{system_overview}.
The system consists of two software modules that run over UDP: a sender-side
program and a receiver-side program. 
\sysname{}~is tightly coupled with Salsify video codec~\cite{fouladi2018salsify}
to dynamically adapt the video offered load to the available network bandwidth.
We chose to integrate \sysname{} with Salsify as it allows to change the
video encoding quality on a per-frame basis.

During a two-way video conferencing session, both users run both the sender and
receiver-side programs. The sender-side program interfaces with the video encoder
and decides how to distribute the encoded frames across available cellular links,
which we refer to as subflows. This scheduling decision is guided by \sysname{}'s
continuous estimate of subflow statistics. In \secref{congestion_control}, we provide
further details on how \sysname{} measures subflow quality. In \secref{scheduler},
we describe our multi-path scheduling scheme.

To quickly adapt to rapid link changes, \sysname{} continuously relays the next frame
target size to the video encoder. This target size represents the total number of bytes
that can be sent across available subflows and delivered within 100 milliseconds.
By capping frame sizes, \sysname{} maximizes perceived video quality while bounding
end-to-end network delays. Details on how target size is computed are provided in \secref{rate_control}.

On the receiver-end, when a new video frame packet is received, \sysname{}'s server-side
program replies with an acknowledgment message (ACKs) and places the packets in a server-side buffer.
To bypass disruptions on the reverse path, we duplicate ACKs across all available interfaces.
Once all packets of the next expected frame are received, \sysname{}'s server-side program
forwards the frame to the decoder for decoding and display. Out-of-order packet deliveries are
common in multipath systems. \sysname{} is not concerned by out-of-order packets of the same frame.
However, out-of-order frames can negatively impact the perceived video quality.
In \secref{ooo_frames}, we provide details on how \sysname{} handles out-of-order video frames.

\subsection{Estimating Per-Subflows Congestion-Window Size}
\label{sec:congestion_control}

By maintaining an estimate of the congestion window (cwnd) size for each
available subflow, \sysname{} can quickly adapt its scheduling decisions
and video encoding quality. The per-subflow congestion window size refers
to the maximum number of bytes that can be sent on each subflow and delivered
within a bounded network delay $\delta$. To estimate the per-subflow cwnd, we
rely on the packets' inter-arrival time as a sign of congestion, similar
to EWMA-Sprout~\cite{winstein2013stochastic} and Salsify~\cite{fouladi2018salsify}.

Upon receiving a new video frame, \sysname{} dissects it into MTU-sized packets
and encapsulates them with header fields that include the frame number,
frame-fragment number, subflow ID, subflow packet sequence number, and inter-frame delay.
The subflow ID allows the \sysname{} receiver to distinguish packets from different
subflows and estimate the per-subflow packet inter-arrival time, which is sent as
part of an acknowledgement message in response to each received packet.
Since video applications capture and display frames at a pre-configured frame rate,
the encoder may pause between frames, leading to a higher perceived inter-arrival time
between the last fragment of one frame and the first fragment of the next frame. To
account for this, we include the inter-frame delay value in the header fields of each video
frame packet. Let $\mathcal{F}$ be the set of available subflows, the inter-arrival
time (iat) of subflow $f$ at the receipt of packet $i$ is thus computed as follows:
\begin{equation}
\label{eq:iat}
\begin{split}
iat^i_f = (t^i_f - t^{i-1}_f) - inter\_frame\_delay
\end{split}
\end{equation}

; where $t^i_f$ is the time packet $i$ is received on subflow $f$.

On the sender-side, each subflow maintains an exponential weighted moving
average of the packets inter-arrival time (ewma\_{iat}) and an estimate of
the propagation delay. The propagation delay estimates allow us to adjust
the end-to-end network delay budget, $\delta$, which has been often overlooked
by existing literature and assumed to be a negligible constant. As we showcase
throughout our manuscript, propagation delay over cellular networks can change
overtime (\eg{} due to cellular hand-offs) and in some cases, can make up a
significant fraction of the end-to-end delay budget for interactive video.
Measuring the per-subflow propagation delay is challenging, therefore we
approximate the propagation delay to be half of the minimum RTT measured in the
last second (min\_RTT). Subsequently, to bound the end-to-end delay on each subflow $f$ to
$\delta$ at each time $t$, \sysname{} caps the total number of in-flight packets
on subflow $f$ to:
\begin{equation}
\label{eq:cwnd}
cwnd^t_f = \frac{\delta - \frac{min\_RTT^t}{2}}{ewma\_iat^t_f}
\end{equation}

\subsection{Adapting video bitrate}
\label{sec:rate_control}

Our choice to integrate \sysname{} with Salsify codec is driven by its ability
to do frame-by-frame optimization. Concretely, when salsify is encoding the
next video frame, it explores multiple encoding qualities, and picks the one that
best matches the available network capacity. This value is referred to as the
next frame target size, which is expressed in bytes. Adjusting the encoding
quality at per-frame level enables Salsify codec to outperform existing video
conferencing applications due to its ability to quickly react to sudden changes
in link capacity.

To estimate the next frame target size, salsify encoder relies on feedback from
the salsify decoder. To enable salsify encoder to leverage all available
cellular link capacities, we intercept the salsify decoder feedback messages
and modify the target-size value to the sum of the aggregate number of bytes
across the available subflows discounted by the number of bytes in flight.
Concretely, we express the next frame target size as follows:

\begin{equation}
\label{eq:framesize}
 target-size^f_t = MTU \times \sum_{f \in \mathcal{F}}((0.8\times cwnd^t_f) - N^t_f)
\end{equation}

; where $N^f_t$ represents the number of packets already in-flight on subflow $f$,
which is obtained by taking the difference between the packets sent and acknowledged
on subflow $f$ at time $t$.

Note that we set the target-size to 0.8 of the cwnd size to account for noise
in the measurement of inter-arrival times and propagation delays.

\subsection{\sysname{} Scheduler}
\label{sec:scheduler}

Given the per-subflow congestion window size, \sysname{}'s multi-path
scheduler decides how to allocate each video frame on the available subflows.
The goal of the scheduler is to minimize the frame's arrival time to
preserve interactivity, while respecting the capacity of each subflow.
Hence, the scheduler must distribute the packets of a video frame such as
to minimize the maximum estimated arrival time of any packet of that frame.
Let $\mathcal{V}$ denote the set of MTU packets that belong to a video frame,
we formulate the problem of scheduling video frames on the available
subflows as follows:\\

\begin{equation}
\label{eq:objective_function}
   Min~Max~\mathcal{M}
\end{equation}

Subject to:

\begin{equation}
\label{eq:constraint_capacity}
  \sum_{v \in \mathcal{V}} x^f_{v,t} <= cwnd^f_t ~~~\forall f \in \mathcal{F}
\end{equation}

\begin{equation}
\label{eq:makespan}
   y^f_t = owd^f_t + (N^f_t + sum_{v \in \mathcal{V}} x^f_{v,t}) * ewma\_iat^f_t ~~~\forall f \in \mathcal{F}
\end{equation}

\begin{equation}
\label{eq:makespan_constraint}
  \mathcal{M} \geq y^f_t ~~~\forall f \in \mathcal{F}
\end{equation}

$\mathcal{M}$ is the objective function, and it indicates the time the
last packet in $\mathcal{V}$ arrives to the receiver. $x^f_{v,t}$ is our
binary decision variable; where  $x^f_{v,t}$ is 1 if fragment $v$ is
allocated to subflow $f$ at time $t$, and 0 otherwise. $y^f_t$ is the
estimated arrival time of the last packet allocated on subflow $f$ at
time $t$. We solve the model as a system of linear equations of the same
form as Equation \ref{eq:makespan}, and then for each subflow, we allocate
the minimum between the outcome of the resolution and its congestion window
size.

\sysname{}'s rate control and congestion control mechanisms are asynchronous,
\ie{} it could happen that by the time the video codec encodes the next
frame with a size that respects the last target size update, the aggregate of
the subflow cwnd size is not enough to accommodate the new encoded frame.
In this case, \sysname{} will fail to schedule the entirety of the new
encoded frame. We therefore equip \sysname{} sender with a queue for 
outstanding frame packets. This queue is sorted in ascending order of 
packet frame deadline. Everytime, \sysname{} receives a new ACK for a
subflow, \sysname{} sender checks to see whether it can send any of the
outstanding packets on that subflow. Before scheduling any of the outstanding
packets, \sysname{} sender first checks to see if the packet frame deadline
has already expired. If the frame deadline has already passed, all
packets of that frame are dropped. \sysname{} then proceeds with
scheduling as many outstanding packets as is available on the subflow within
the limits of its cwnd.

\subsection{Handling Out of Order Frames}
\label{sec:ooo_frames}

\sysname{} receiver maintains an index of the next expected frame, and
a per-frame buffer. At the start of a video conferencing session, the next expected
frame is set to 0.
Each time a new frame packet is received, \sysname{} receiver checks
whether a buffer for this frame already exists. If not,
it starts a new buffer for the frame, and appends the frame in the
buffer at the index corresponding to the fragment number indicated in the
packet header field.

Once all packets of the next expected frame are received, \sysname{}
forwards all the packets of the frame to salsify decoder to be
decoded and displayed. At the receipt of a new frame, salsify decoder replies
with salsify feedback messages, that are forwarded to salsify encoder via
\sysname{} sender. Salsify feedback messages enables
salsify encoder to know the $\it{state}$ at the decoder side. This state
information is critical for salsify encoder, as it allows it to know
which state it can use for the next frame encoding. If no salsify feedback is
received, salsify encoder goes into loss recovery mode, and rolls back the
encoding state to the last acknowledged state. Hence, \sysname{} receiver
must carefully relay frames to the salsify receiver, since out of order frame
delivery will falsely put salsify encoder in loss recovery state.

Let $s_i$ be the next expected frame. Consider the case where \sysname{}
receiver has received only a subset of the packets of frame $s_i$, and that
$s_i$ deadline has expired. \sysname{} waits until the deadline of $s_{i+1}$ - $\omega$ before
it forwards whatever packets of frame $s_i$ it has in the buffer; where
$\omega$ represents time to decode the frame. The
rationale behind this is, that given that salsify decoder depends on the state
of frame $s_{i}$ to decode frame $s_{i+1}$, \sysname{} holds-off from forwarding
incomplete frames until the deadline of the next frame. This provides
a grace period to receive the outstanding packets of frame $s_{i}$ needed to
update the decoder state and properly decode frame $s_{i+1}$.

\subsection{Adaptive Probing}

\sysname{} relies on the ewma\_iat for its scheduling, congestion-control,
and rate-control schemes. Hence, receiving continuous updates of the iat
on each subflow is critical to enable \sysname{} to timely adapt to changes
along the available cellular links. Consider the case where a subflow has
no in-flight packets, and its subflow cwnd drops to zero. To \sysname{}
sender, the subflow will appear unusable indefinitely.

Existing work~\cite{fouladi2018salsify} addresses this situation using
periodic probing every ~100 milliseconds. However, our experiments have shown
that in the event of poor link conditions, periodic probing can lead
to queue build-ups further delaying the time to recovery.
Instead, \sysname{} employs an adaptive probing strategy. Let
$t_f$ be the time when the last packet was sent on subflow $f$. \sysname{}
sets the subflow probing rate to
\begin{equation}
\tau^f_t = min(1,2\times{ewma\_iat}^f_t)
\end{equation}
; where $\tau^f_t$ indicates how
often we can push a pair of packets on subflow $f$ without violating the
link capacity. Subsequently, at the receipt of a new frame, \sysname{} checks
if no pair of packets have been scheduled on subflow $f$ for more than
$\tau^f_t$ seconds, \sysname{} duplicates two packets of
the new frame and allocate them to this subflow. This adaptive probing scheme
allows \sysname{} to continuously update the subflow statistics
without any queue build-ups when the cellular link capacity is low.

\section{Evaluation TestBed}
\label{sec:cellular_emulator}

To evaluate \sysname{}'s performance, we employ trace-driven emulation, which
allows us to capture real-world network conditions and reproduce them
repeatedly for fair comparisons with state-of-the-art video conferencing solutions.

\subsection{Limitations of existing trace-driven emulator}
\label{sec:emulator_limitations}
Several trace-driven emulators are available~\cite{netravali2015mahimahi,
winstein2013stochastic,mishra2021nemfi}, each tailored to a specific application.
However, they all share a common framework: a recording tool to capture network
traces and a replay tool that emulates the captured network conditions using the
recorded traces. One emulator, called cellsim~\cite{winstein2013stochastic},
is specifically designed to record and replay cellular network conditions.
Cellsim assumes that propagation delay is constant, so it focuses on
capturing and reproducing queuing delays experienced over cellular links.
To do this, the cellsim record tool saturates
the network pipe in both directions and records the arrival time of each packet.
The result is a trace of  $\it{delivery~opportunities}$, which represent the relative
time each packet will spend in the queue during replay before being delivered to the
other end. Before being placed in the replay queue, cellsim delays each packet for a
fixed amount of time to simulate propagation delay.

One limitation of Cellsim is that it assumes constant propagation delay.
Previous studies and our own measurements show that
propagation delays can vary due to Internet route changes and hand-offs.
To measure the rate and amplitude of propagation delay changes in mobile
scenarios\footnote{The findings in \cite{pucha2007understanding}
pertained only to static scenarios, and demonstated that propagation delay changes
can occur even in such setups, albeit with a low rate of change that may take days to
manifest}, we developed a simple trace collection tool
that records one-way delays (OWD) in both directions between a client machine
connected to a 4G router and a well-connected server. The client and server
machines are connected to a GPS (Navisys GR-701W) and are time-synchronized
using NTP, providing a time synchronization accuracy of 1 ms. To avoid queuing
delays that could impact OWD measurements, we sent a small probe (50 bytes) every
100 milliseconds from each endpoint to the other. At the receiving end, we logged
the measured OWD for each probe. We collected data while driving a car in
a large metropolitan area with four different cellular network operators, conducting
multiple 15-minute measurements for each operator. In total, we collected two hours
worth of one-way delay measurements.

\begin{figure}[t!]
    \begin{subfigure}[t]{0.5\linewidth}
        \centering
        \includegraphics[height=1.2in]{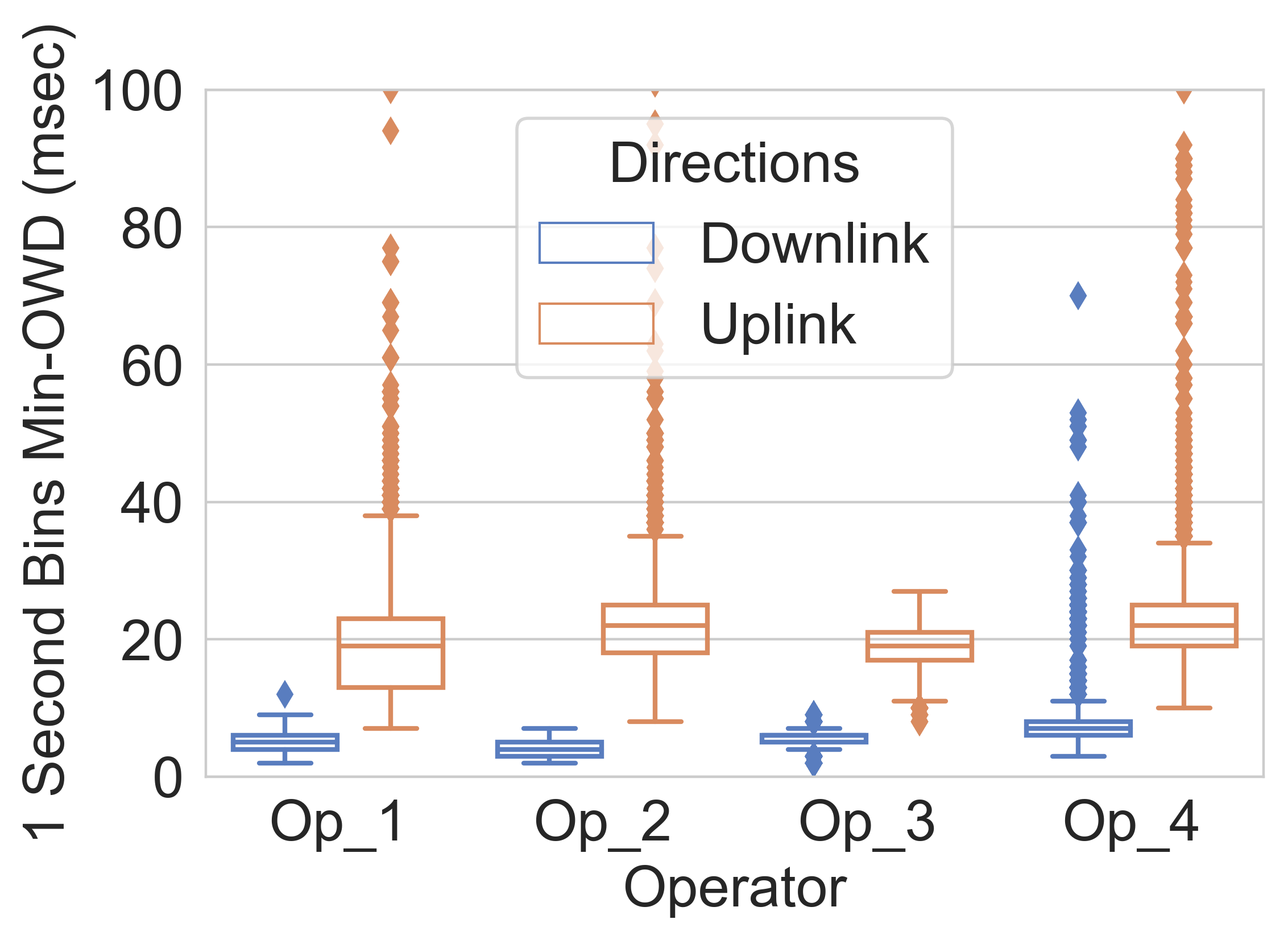}
        \caption{Min-OWD 1-Sec Bin}\label{fig:per_bin_owd}
    \end{subfigure}%
    \begin{subfigure}[t]{0.5\linewidth}
        \centering
        \includegraphics[height=1.2in]{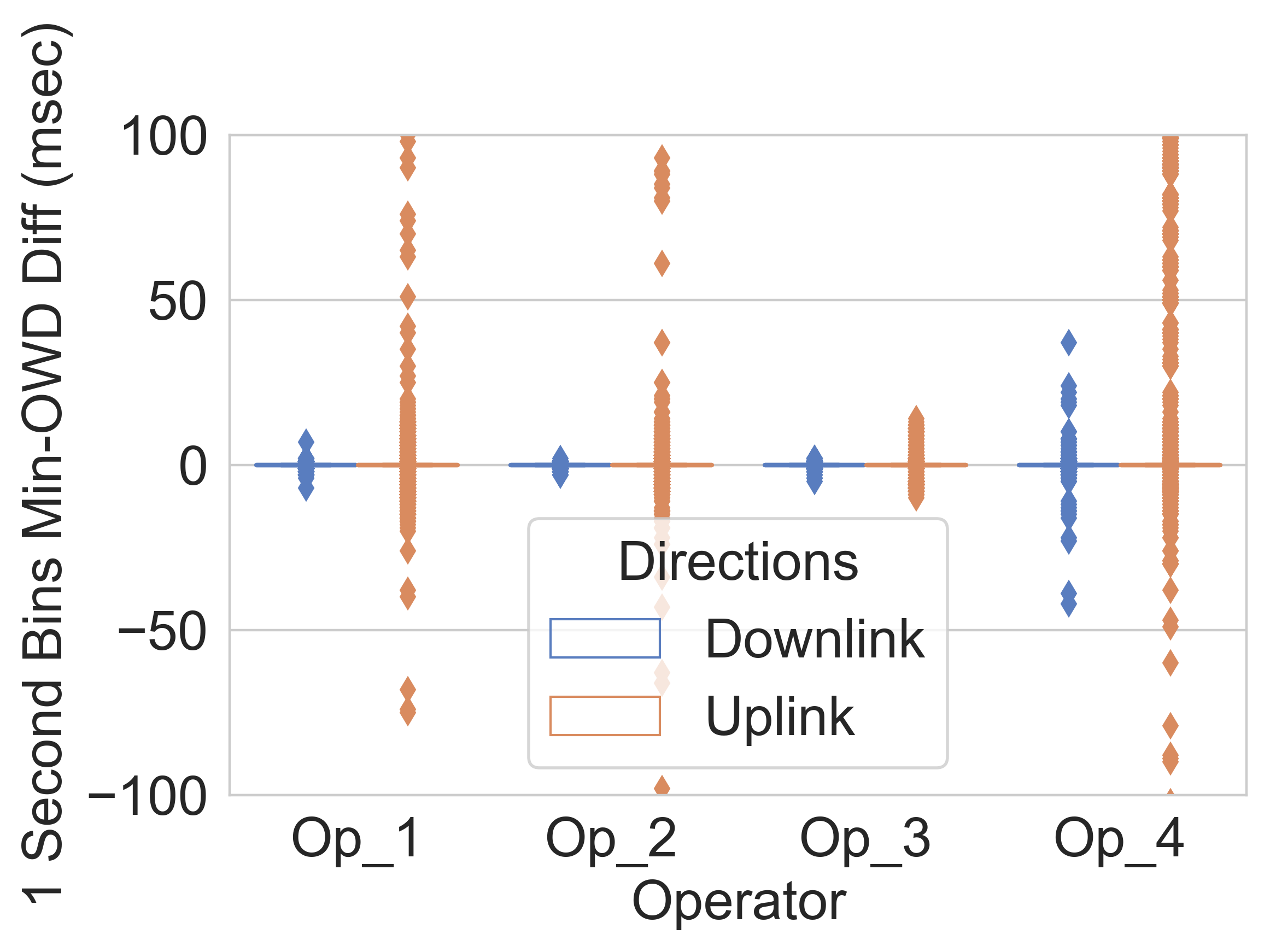}
        \caption{Min-OWD Change for Consecutive Bins}\label{fig:bin_owd_diff}
    \end{subfigure}
    \begin{subfigure}[t]{\linewidth}
        \centering
        \includegraphics[height=1.2in]{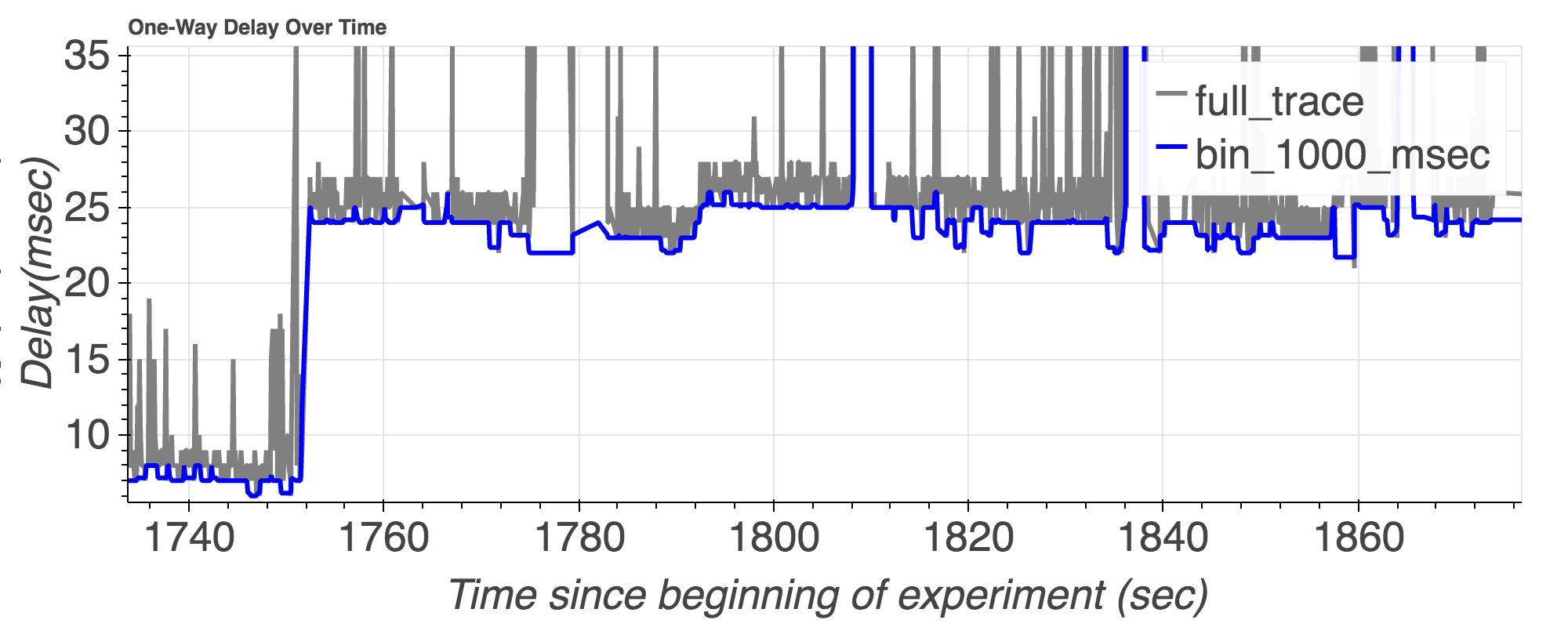}
        \caption{Downlink OWD timeseries for operator 4}\label{fig:sfr_cpd}
    \end{subfigure}
    \caption{Measurements of owd changes across different cellular operators.}
\label{fig:owd_variations}
\end{figure}

To illustrate our findings, we plot the minimum OWD obtained in every
one-second bin for each operator in \figref{per_bin_owd}. Our results show that
while the average behavior of propagation delays is generally stable over time,
the OWD on the uplink and downlink are not symmetrical. We also find
that propgation delay changes can happen within a 15-minute session as indicated
by the outlier data points. Further, to understand the amplitude of change in
OWD, we plot the difference in minimum OWD obtained in consecutive
one-second bins (as illustrated in \figref{bin_owd_diff}.  Our results show that
the amplitude of change between two consecutive one-second bins can go up to
several tens of milliseconds. Finally, by examining the OWD trace for a
single operator (\figref{sfr_cpd}), we observe that the change in OWD is
not necessarily abrupt, but can indeed last for several consecutive time bins.

It's important to note that although cellsim does not explicitly capture
propagation delay changes,
the effect of internet route changes will be inadverently captured during the
record. However, to accurately reproduce this behavior, the application must
send at full-throttle during replay. In this case all delivery
opportunities recorded will be replayed. Subsequently, the data packet sent
after the propagation delay change occurs will make all subsequent packets
queue up behind it and wait for the same amount. However, interactive video
applications follow an on-off sending pattern, where the application pauses
between frames. In this case, the delivery
opportunity that captures the propagation delay changes will only delay the
other packets sent as part of the same burst, and its impact will then be
dilluted, or even completely disappear if the amplitude of propagation delay
change is smaller than the pause duration.

\begin{figure}
    \begin{subfigure}[t]{0.5\linewidth}
         \centering
         \includegraphics[height=1.2in]{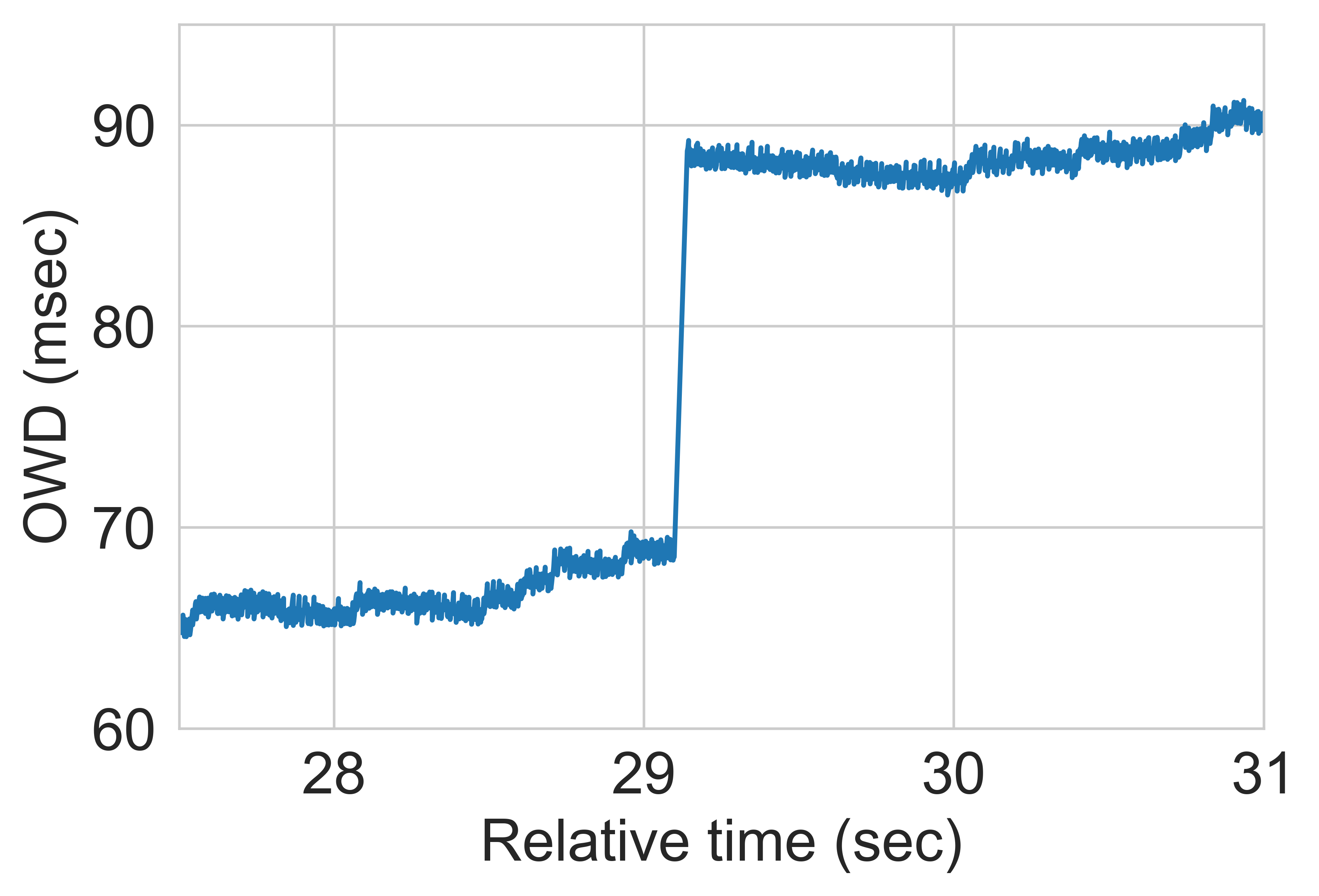}
         \caption{OWD with a full-throttle sending rate}\label{fig:per_bin_rtt}
     \end{subfigure}%
     \begin{subfigure}[t]{0.5\linewidth}
         \centering
         \includegraphics[height=1.2in]{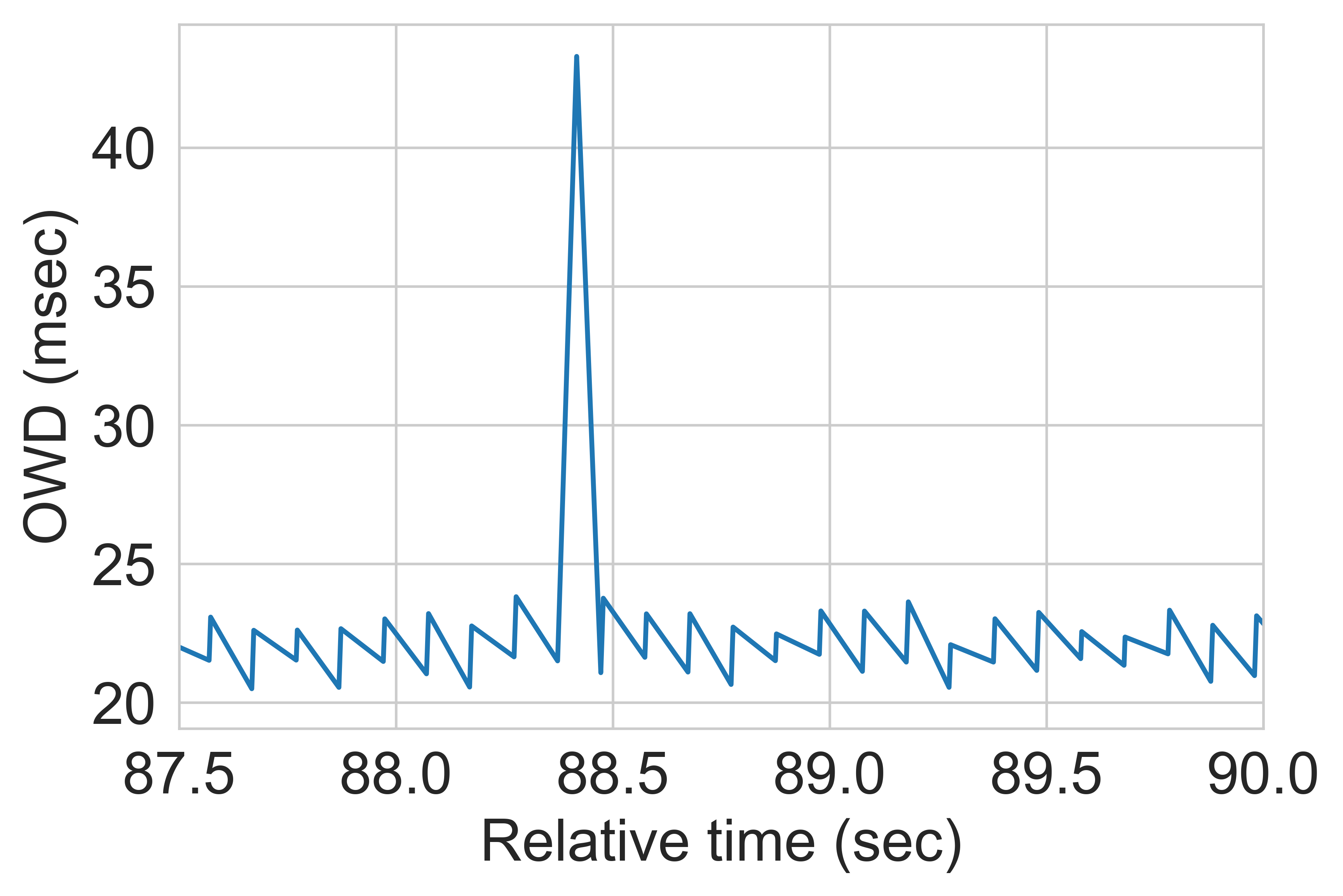}
         \caption{OWD with an on-off sending rate}\label{fig:bin_rtt_diff}
     \end{subfigure}
     \caption{Measured owd during replay (full-throttle vs on-off).}\label{fig:mpshell_replay}
 \end{figure}
To better illustrate this, we ran the record tool of cellsim between a client and a
server machine over a cable link; and we limit the bandwidth of the link to
six Mbps. We begin the experiment by setting the round-trip propagation delay
to 20 milliseconds and then increase it to 60 milliseconds to simulate a
propogation delay change. Figure \ref{fig:mpshell_replay} illustrates the
measured OWD when we replay with a bulk-transfer type of
application versus an on-off application (here we send a burst of three
MTUs and pause for 100 msec akin to a sending rate of 360 kbps at ten fps).
Clearly, the impact of the propagation delays is accurately
reproduced when the application sends at bulk-transfer capacity; whereas with
the on-off application, only a single packet incurs an increase in OWD, while
the subsequent data packet are not affected by the change.

To address the limitations of cellsim, we introduce \emuname{}, an extension of cellsim that
captures propagation delays during the record phase and can faithfully reproduce them
during replay regardless of the application sending pattern.

\subsection{\emuname{}:~Cellular Networks Emulator}
\label{sec:emulator_description}

Given the existing limitations of Cellsim, we extend Cellsim to propose
\emuname{}: a multi-path emulator for cellular networks. 

\begin{figure}
    \begin{subfigure}[b]{0.45\textwidth}
        \centering
        \includegraphics[width=1\linewidth]{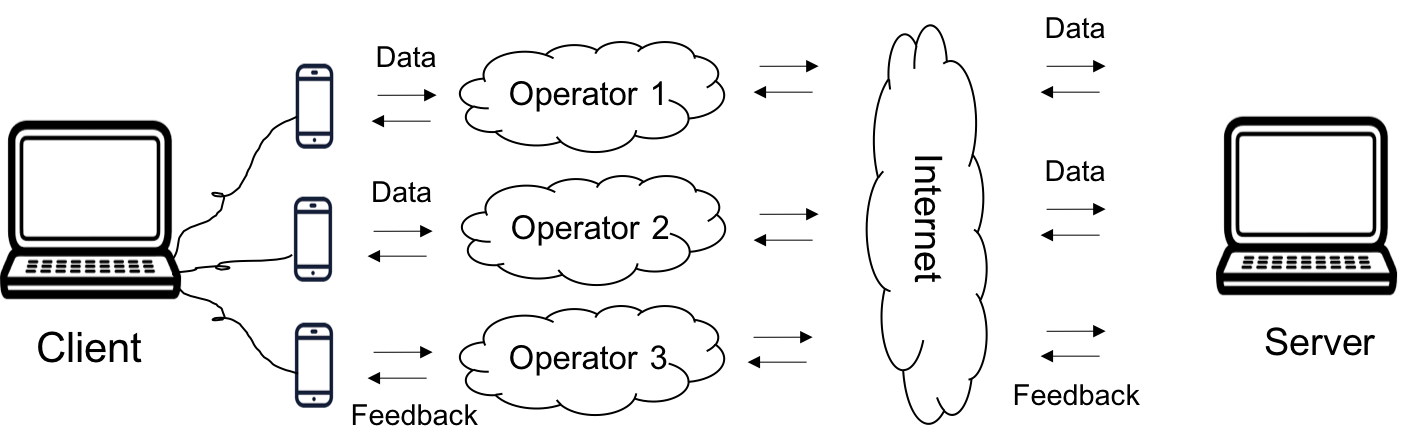}
        \caption{\emuname{} Record Block Diagram}\label{fig:TPE_record}
    \end{subfigure}%
    \\
    \begin{subfigure}[b]{0.45\textwidth}
        \centering
        \includegraphics[width=1\linewidth]{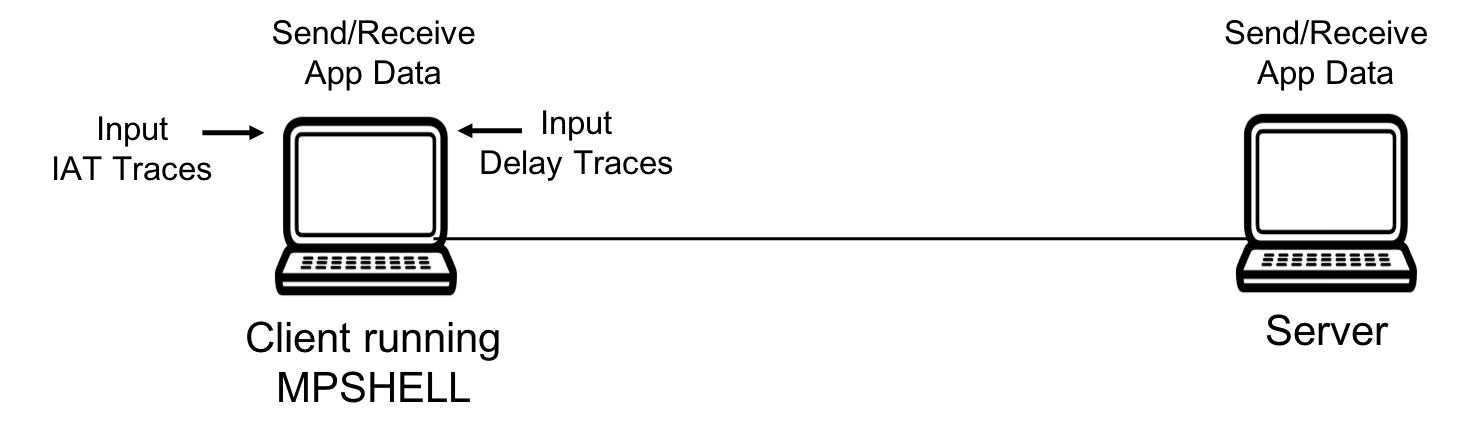}
        \caption{\emuname{} Replay Block Diagram}\label{fig:TPE_replay}
    \end{subfigure}
    \caption{Block Diagram of \emuname{} for Cellular Networks.}\label{fig:TPE_emulator}
\end{figure}

The key novelties
of \emuname{} are: 1) It directly measures propagation delays in-addition to
bottleneck link capacity of cellular networks, and 2) it supports
applications with on-off sending patterns. In the sequel, we detail
how \emuname{} works. Figure \ref{fig:TPE_emulator} illustrates the block
diagram for \emuname{}'s record and replay. From a setup point of
view, \emuname{}'s record (illustrated in Figure \ref{fig:TPE_record}) is
very similar to that of cellsim, with the exception that we run the
record on two cellular interfaces instead of just one, since we are
interested in capturing network variabilities that occur on two cellular
networks at the same time.
Moreover, \emuname{} records both
variations in bottlneck link capacity and propagation delays. The main challenge is that
measuring bottleneck link capacity and propagation delays simultaneously is not
possible \cite{cardwell2016bbr}. To measure bottleneck link capacity, we
need to send enough data to fill the pipe, which will create a queue at the
bottleneck link and subsequently increase the delay. 

Hence, to record cellular network variations, we need to alternate between
saturating the pipe to measure the bottleneck link capacity, and sending
small well-spaced probes to capture propagation delays. To achieve this, we
need to identify how often to switch between these two modes; which in
essence depends on how often propagation delay changes happen to make sure
that we capture them when they occur.\\

\par{\textit{How often do we need to measure propagation delays?}~To answer this question,
we look at how often significant propagation delay changes happen across our
in-car measurements. Similarly to \secref{emulator_limitations},
we take the minimum OWD for each one second of each experiment, then we apply
bayesian change point detection method on each trace. The goal is to detect
significant bins of change, and thereby deduce the size of the bins where the
OWD remains more or less stable. We consider change to be significant if the
average OWD of two consecutive bins vary (increase/decrease) by at least two
milliseconds, which is a conservative bound.

\begin{figure}
  \centering
  \includegraphics[width=\linewidth]{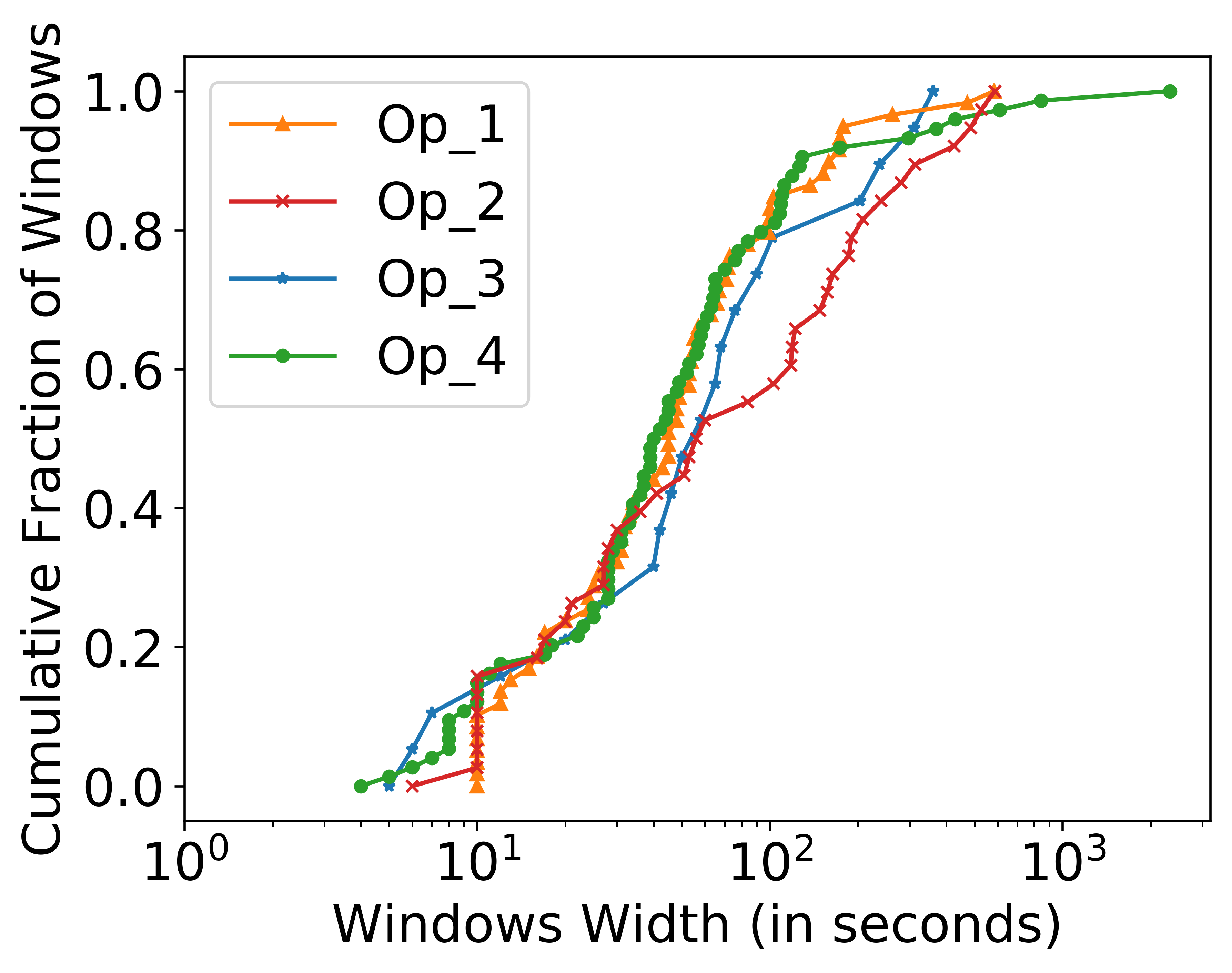}
  \caption{CDF persistent owd bin size }\label{fig:cpd_owd}
\end{figure}

\figref{cpd_owd} illustrates a CDFs of all the bin sizes obtained for
each operator; where each pair of consecutive bins delimits a change point.
We observe that almost all bins are at least ten seconds long; with the
minimum bin size value obtained across all experiments is ~5 seconds. Hence,
we adopt a frequency switch value of 5 seconds throughout all of our
measurements.

 We now describe how \emuname{} achieves high fidelity record
of cellular network conditions and a replay that is suitable for
applications with either bulk-transfer or on-off sending pattern.

\par{\textit{\emuname{}-Record:} \emuname{}'s record consists of two-phases
that repeat iteratively in
cycles:
\begin{itemize}
\item Phase 1 - Saturation: Saturation mode consists of running the
record routine of cellsim as-is \cite{winstein2013stochastic} by
sending at full-throttle to saturate the pipe. We
run the saturation routine for
5 seconds and record the arrival time of each data packet. Once the
saturation phase is over, the delay measurement routine begins.
However, during saturation phase a queue may build up;
and hence we cannot switch to measuring the propagation delays right away
since the bloated queue will affect the one-way delay measurements. Thus
before measuring propagation delays, \emuname{} goes into a drainage phase.
The end of the drainage phase is signaled by the receipt of an ACK for the
last data packet sent in saturation mode or when a timeout period elapses.

\item Phase 2 - Delay Measurement: Delay measurement phase consists of
sending small size probes every 100 milliseconds and logging the one-way
delay of each at the receiver end. Similarly to the saturation phase, we run the
delay measurement phase for up to 5 seconds. The effective duration of the
delay measurement phase depends on the time spent waiting for the queue to
drain, and can be anywhere between 2 to 5 seconds.
\end{itemize}

\begin{figure}
  \centering
  \includegraphics[width=\linewidth]{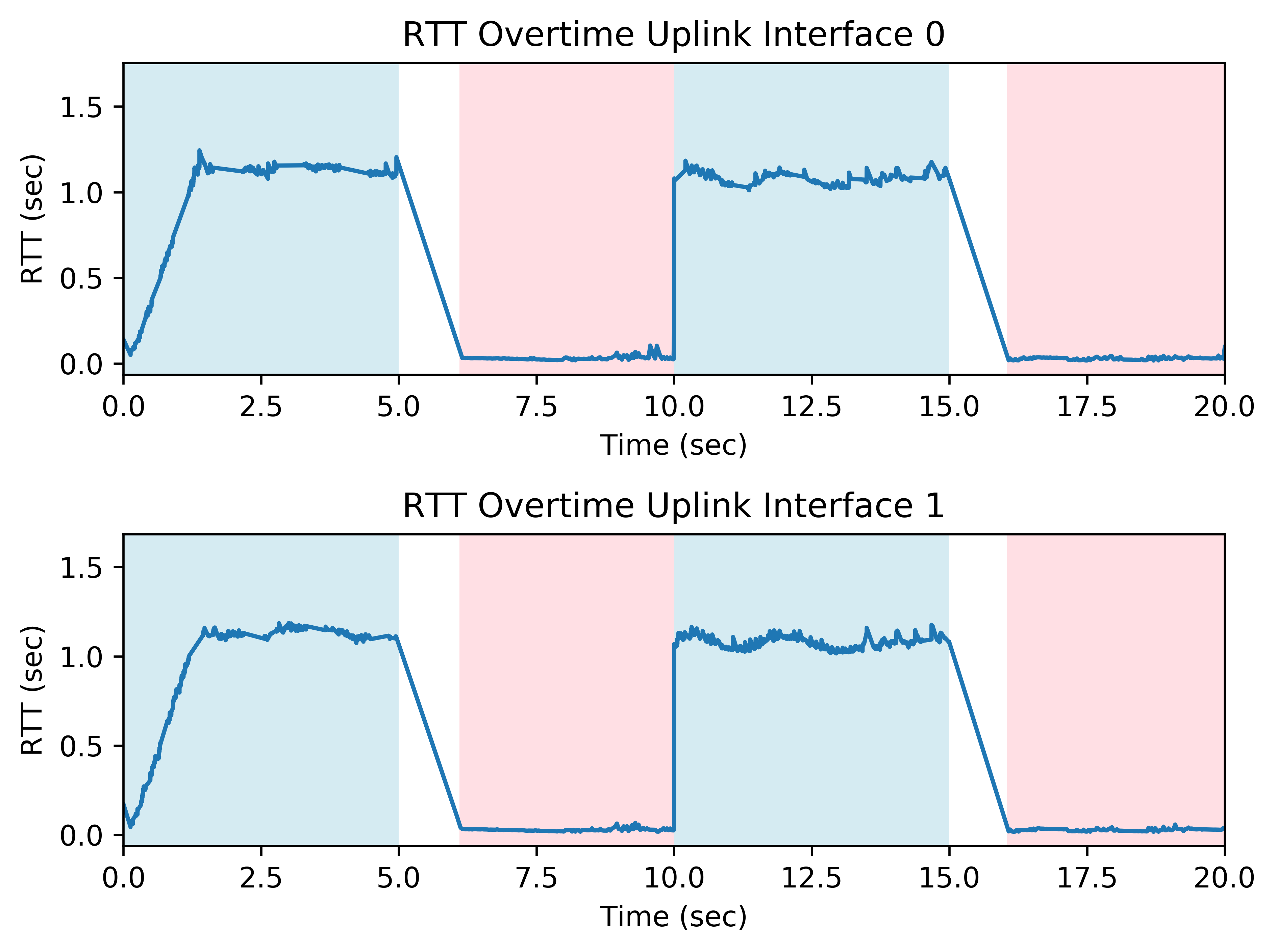}
  \caption{Example of two cycles captured with the Two-Phases Saturator}\label{fig:2p_sat}
\end{figure}

Figure \ref{fig:2p_sat} shows an example of a trace recorded with the
\emuname{} record tool. It illustrates two cycles recorded over
two interfaces simultaneously; where each cycle consists of a saturation
phase (blue window) followed by drainage period (white window) and then the
delay measurement phase (pink window). The end of a delay measurement phase
signals the end of a cycle.}

\par{\textit{\emuname{}-Replay:}~ For the replay, we use MpShell, the
multi-path version of Cellsim \cite{deng2014wifi}, and we modify it to take
as input for each interface in each direction a trace of packet Inter-Arrival
Times (IAT) (similar to Cellsim and the original MpShell) and a propagation
delays trace. Recall that the packets inter-arrival times and propagation
delay measurements were conducted sequentially in cycles. \emuname{}'s replay
overlay the IAT and propagation delay measurements captured in each cycle as
if they were recorded at the same time. Hence, the duration of the replay
traces is half of the actual record time. For each cycle, \emuname{}'s replay
takes the minimum propagation delay value recorded in each cycle, and
apply this value to all the packets sent during that cycle. The packets are
then released from the queue according to the recorded IAT trace. \emuname{}
adapts the propagation delay value used at the beginning of each cycle. If
the experiment time is larger than the replay trace duration, \emuname{}
loops back to the beginning of the trace.}

\section{Numerical Evaluation}
\label{sec:evaluation}

In this section, we evaluate \sysname{}'s ability to improve the perceived
quality of video conferencing sessions. We use CellNem to record network
traces using six different pairs of cellular network operators in both indoor
and outdoor settings. The indoor trace collection is conducted in a stationary
setup from an apartment building in a larger metropolitan city, while the
outdoor trace collection is conducted in a mobile setting while driving
a car in a large metropolitan city. \tabref{dataset} provides a summary of the
network traces captured.

\begin{table*}[!htbp]
\begin{tabular}{|l|c|c|c|c|c|}
\hline
\multicolumn{1}{|c|}{\textbf{Location}} & \textbf{Mode} & \textbf{\#Traces} & \textbf{\#Operators} & \textbf{\begin{tabular}[c]{@{}c@{}}Min/Max Average Bandwidth \\ Uplink (Mbps)\end{tabular}} & \textbf{\begin{tabular}[c]{@{}c@{}}Min/Max Average Bandwidth \\ Downlink (Mbps)\end{tabular}} \\ \hline
Indoor  & Stationary  & 10  & 4 & {[}0.82 - 40.85{]}  & {[}0.33 - 142.49{]}  \\ \hline
Outdoor & Mobile & 11  & 4  & {[}1.26 - 41.85{]}  & {[}8.32 - 73.65{]}   \\ \hline
\end{tabular}
  \caption{Cellular network traces collected in different locations of a metropolitan area}\label{tab:dataset}
\end{table*}

\begin{figure*}[htb]
    \begin{subfigure}[b]{0.33\textwidth}
        \centering
        \includegraphics[width=1\linewidth]{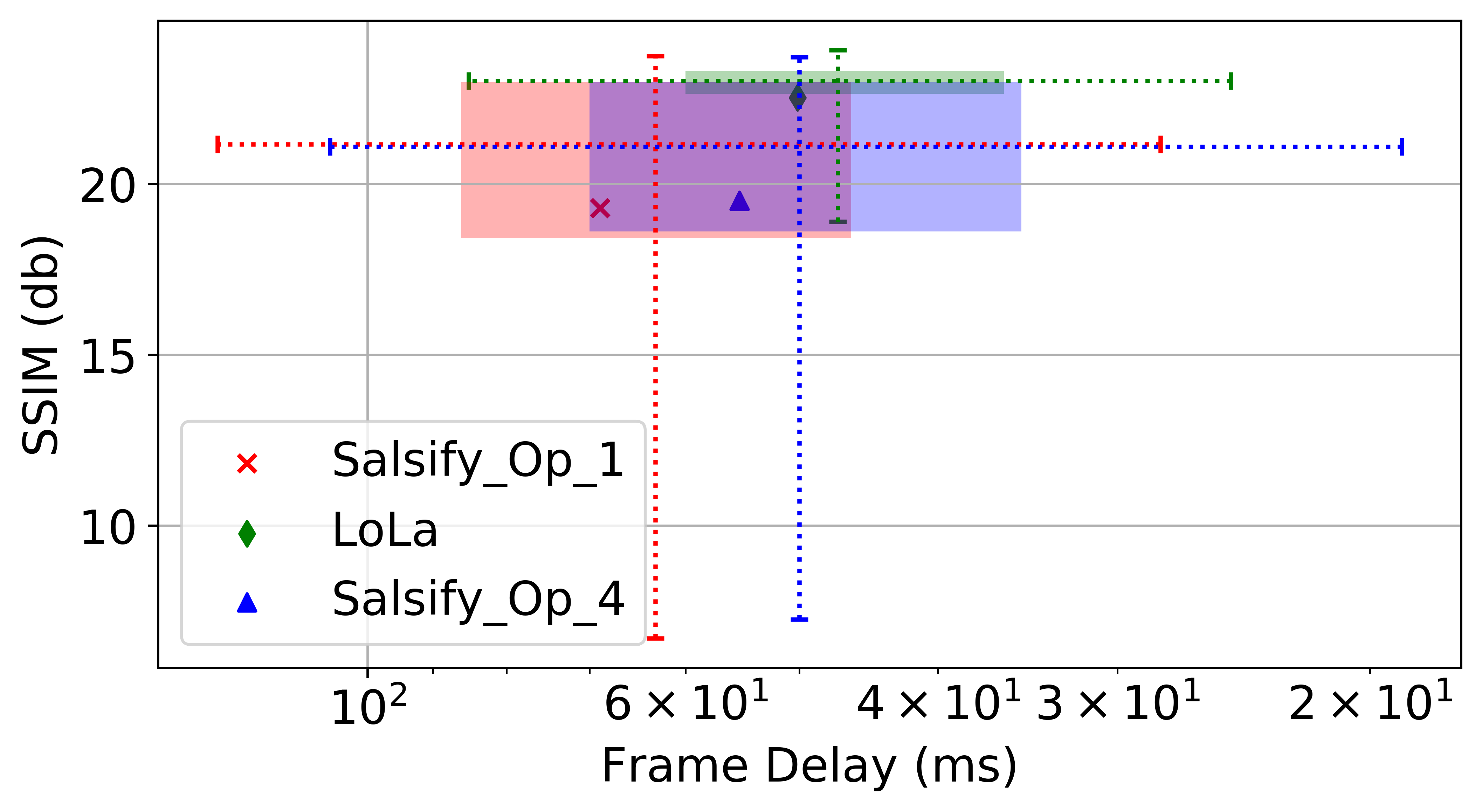}
    \end{subfigure}
    \begin{subfigure}[b]{0.33\textwidth}
        \centering
        \includegraphics[width=1\linewidth]{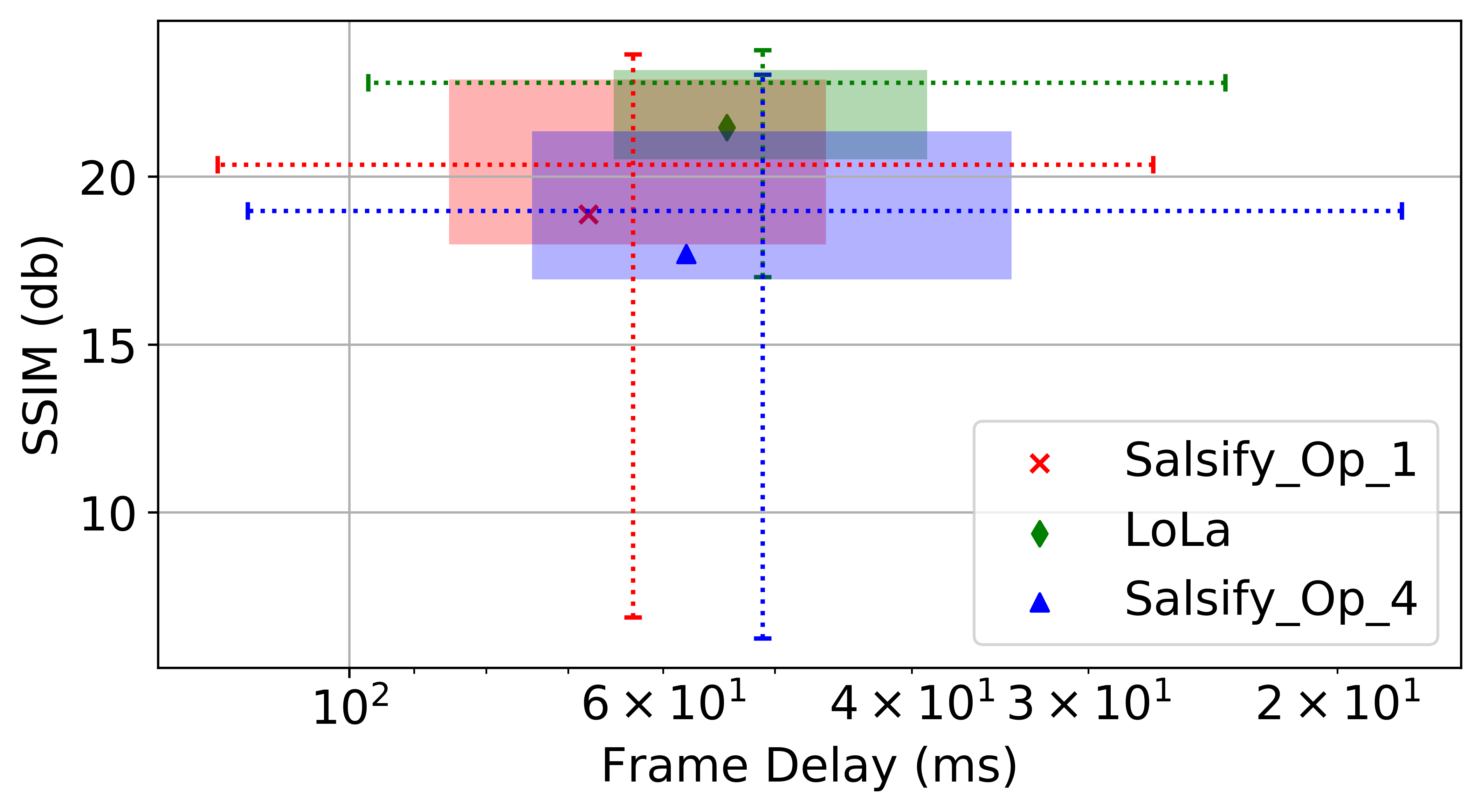}
    \end{subfigure}
    \begin{subfigure}[b]{0.33\textwidth}
        \centering
        \includegraphics[width=1\linewidth]{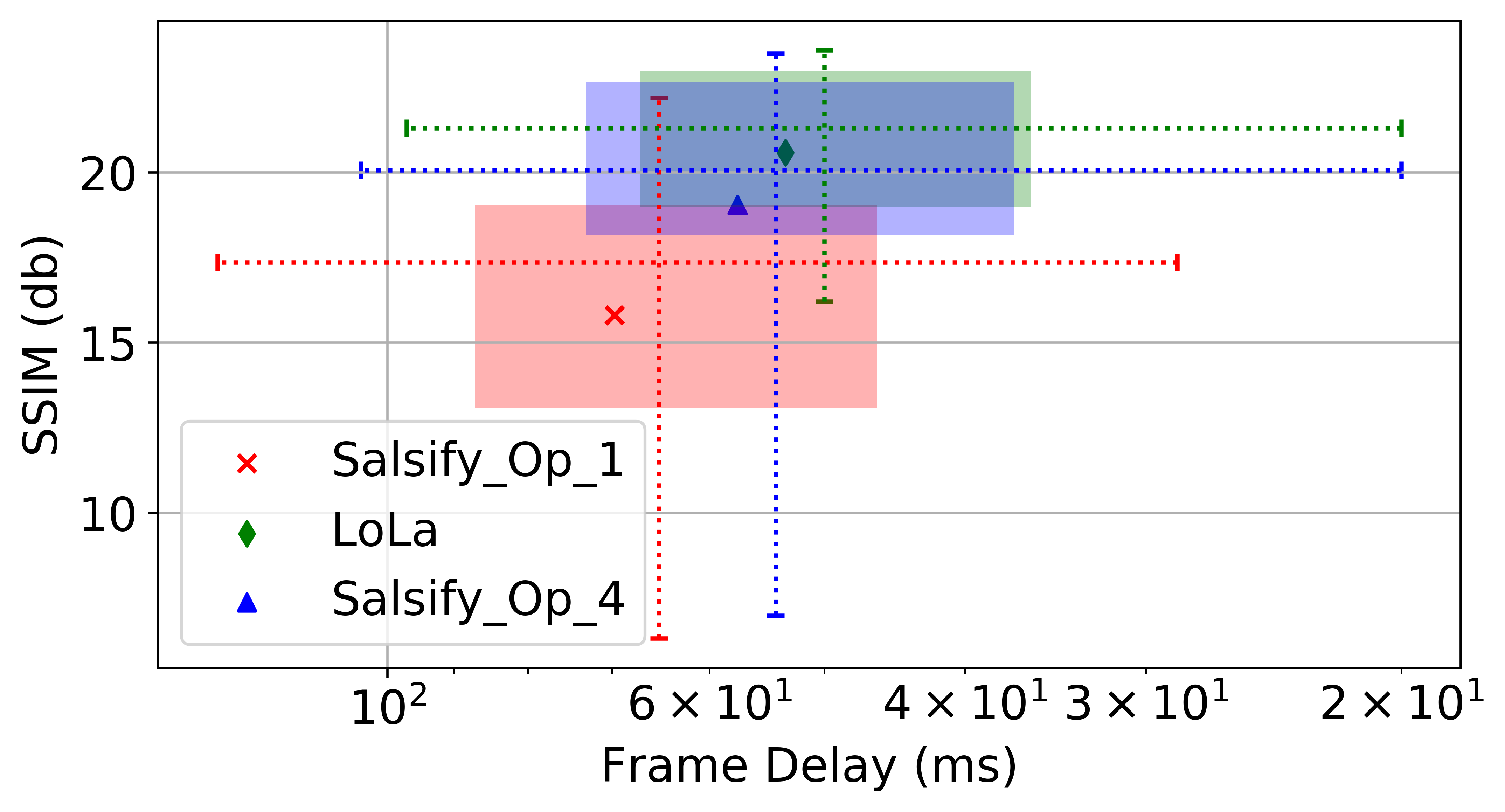}
    \end{subfigure}
    \begin{subfigure}[b]{0.33\textwidth}
        \centering
        \includegraphics[width=1\linewidth]{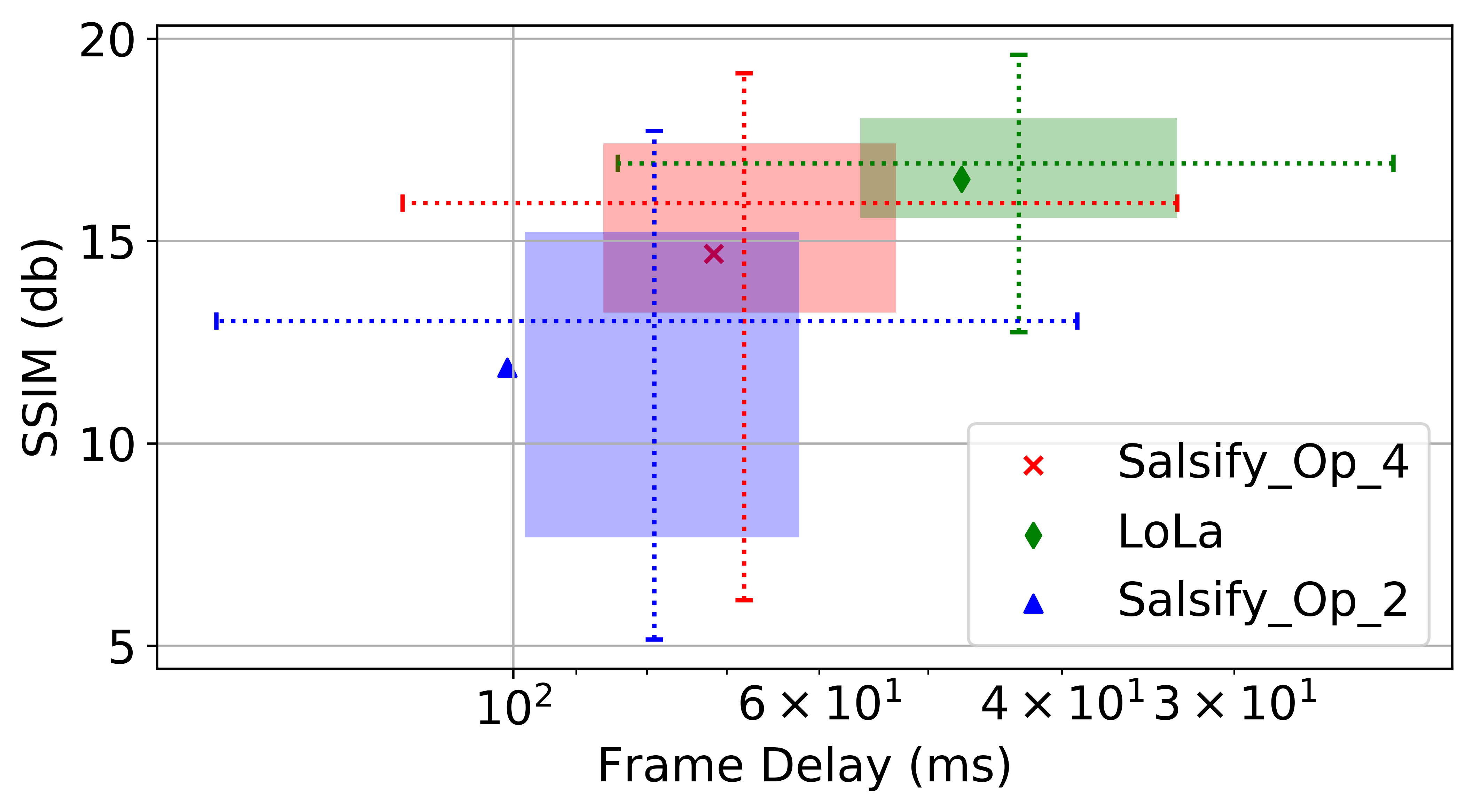}
    \end{subfigure}
    \begin{subfigure}[b]{0.33\textwidth}
        \centering
        \includegraphics[width=1\linewidth]{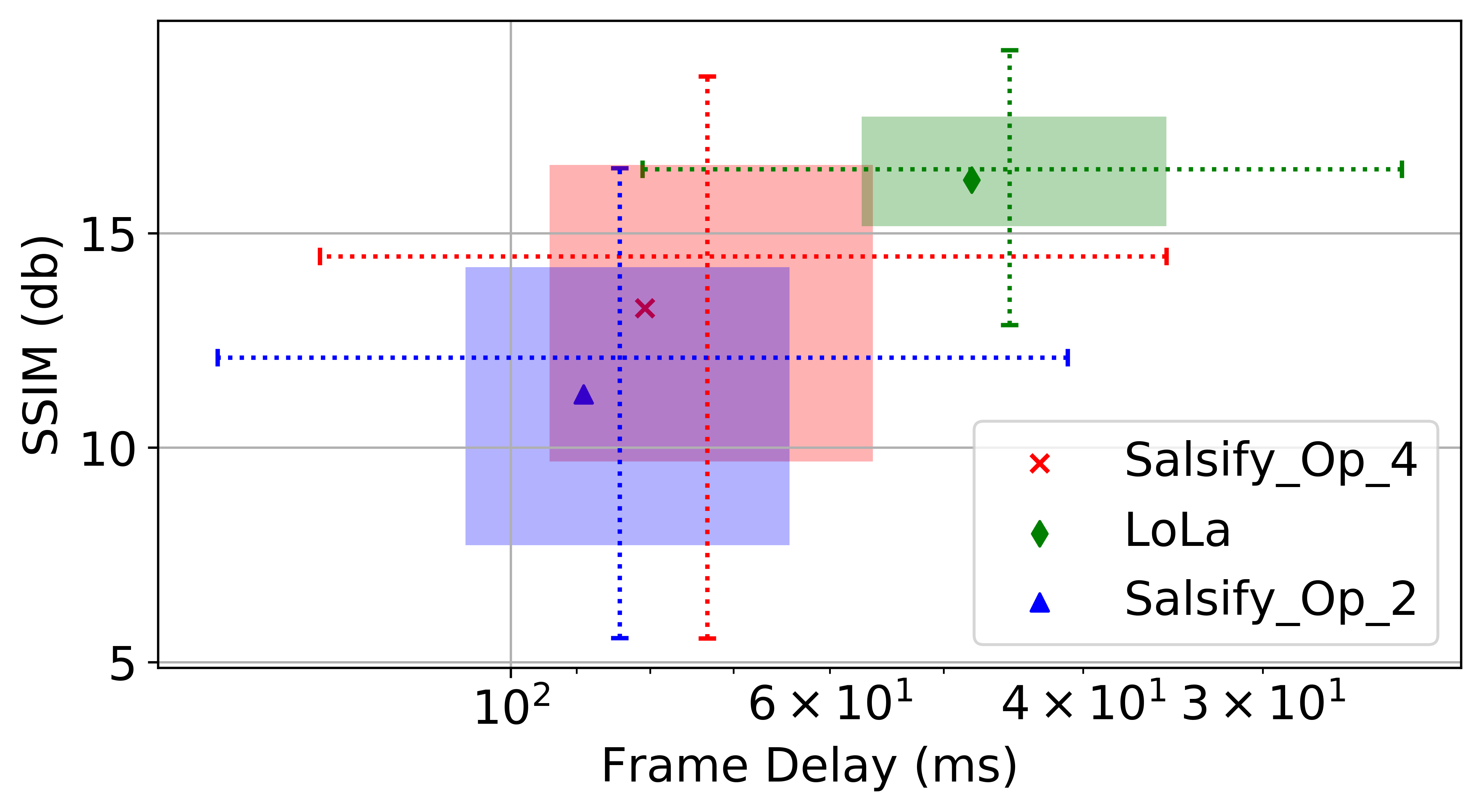}
    \end{subfigure}
    \begin{subfigure}[b]{0.33\textwidth}
        \centering
        \includegraphics[width=1\linewidth]{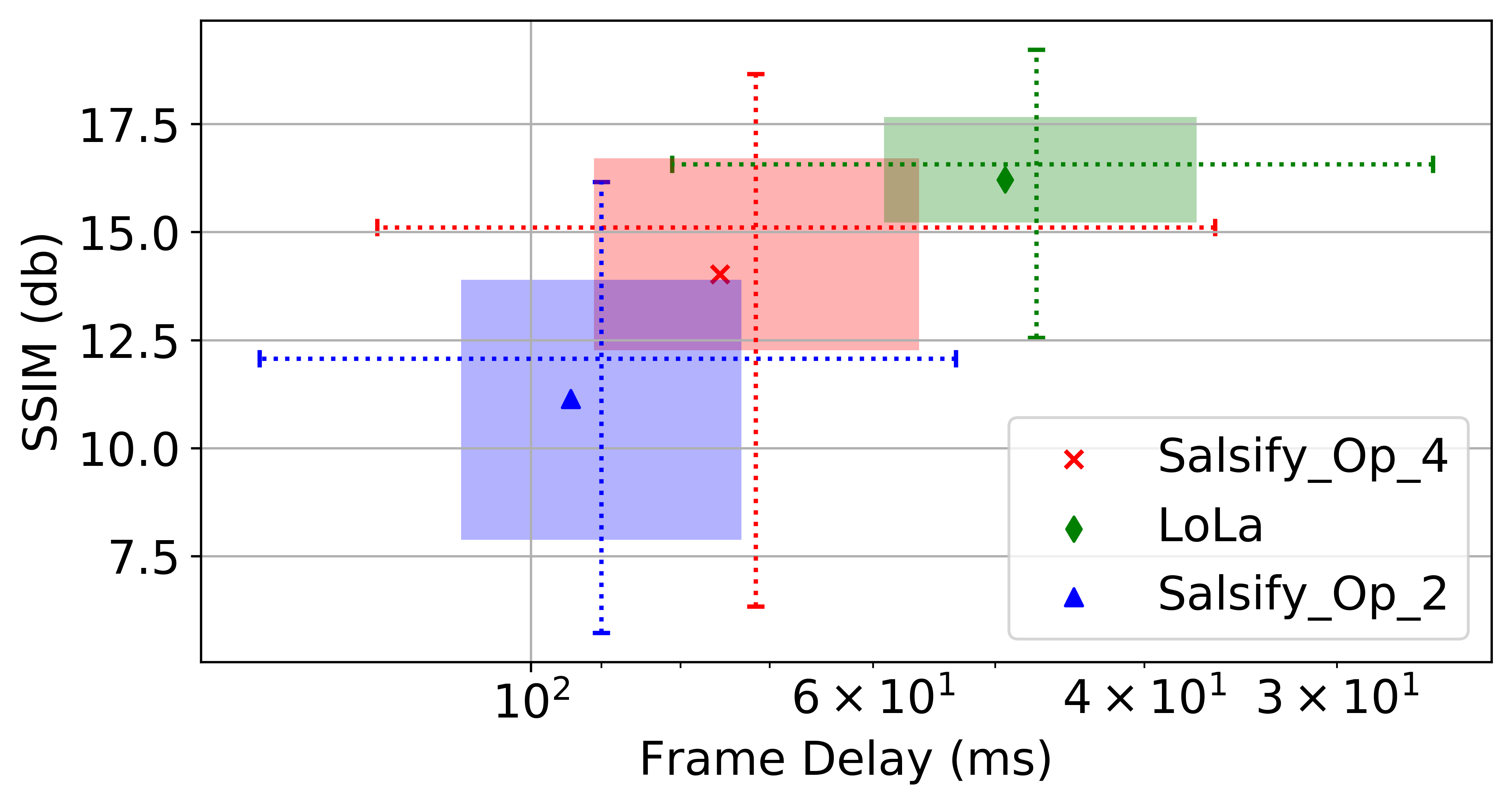}
    \end{subfigure}
    \caption{\sysname{} vs. Salsify in Indoor Stationary Setting}\label{fig:stationary_results}
\end{figure*}
\begin{figure*}[htb]
    \begin{subfigure}[b]{0.33\textwidth}
        \centering
        \includegraphics[width=1\linewidth]{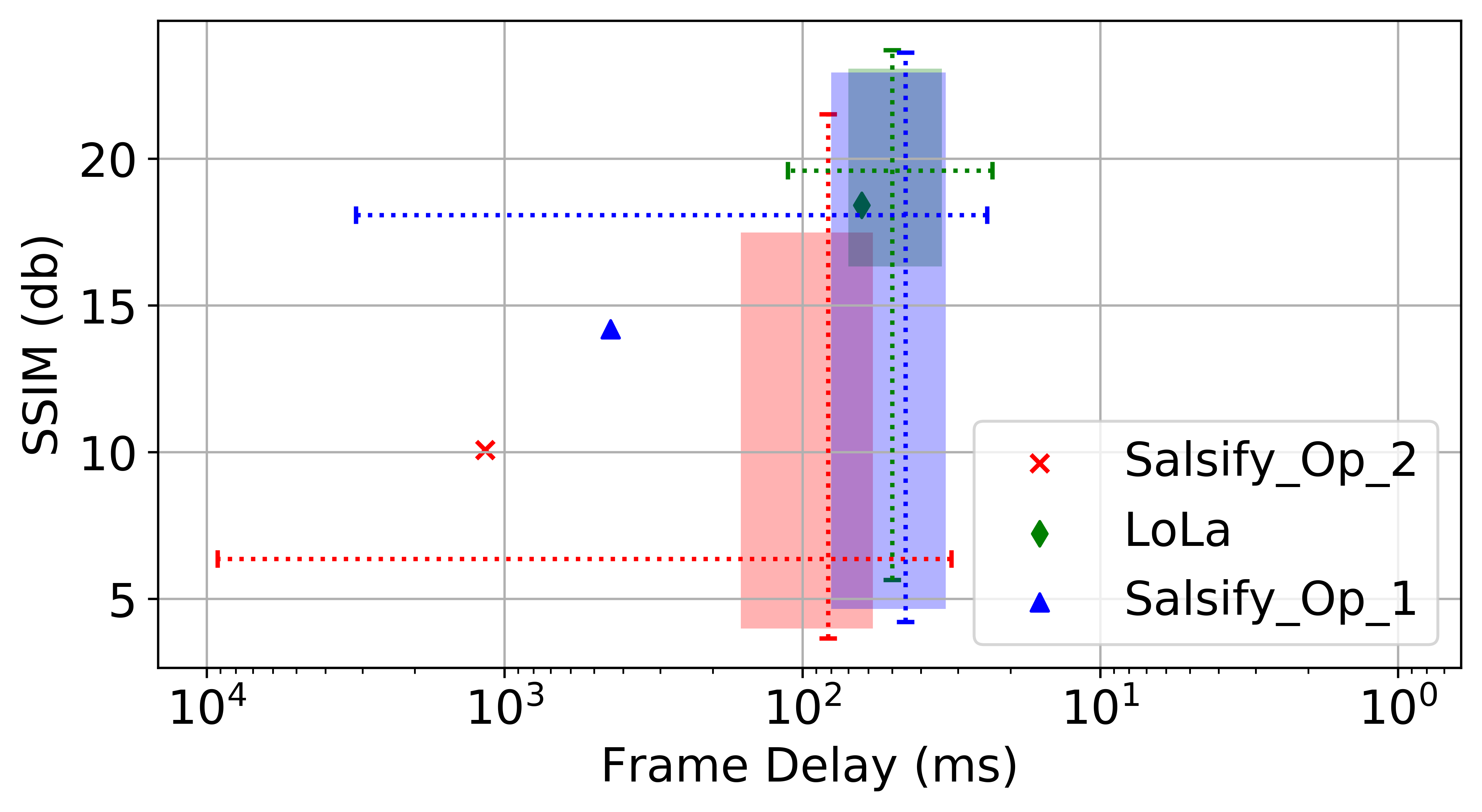}
    \end{subfigure}
    \begin{subfigure}[b]{0.33\textwidth}
        \centering
        \includegraphics[width=1\linewidth]{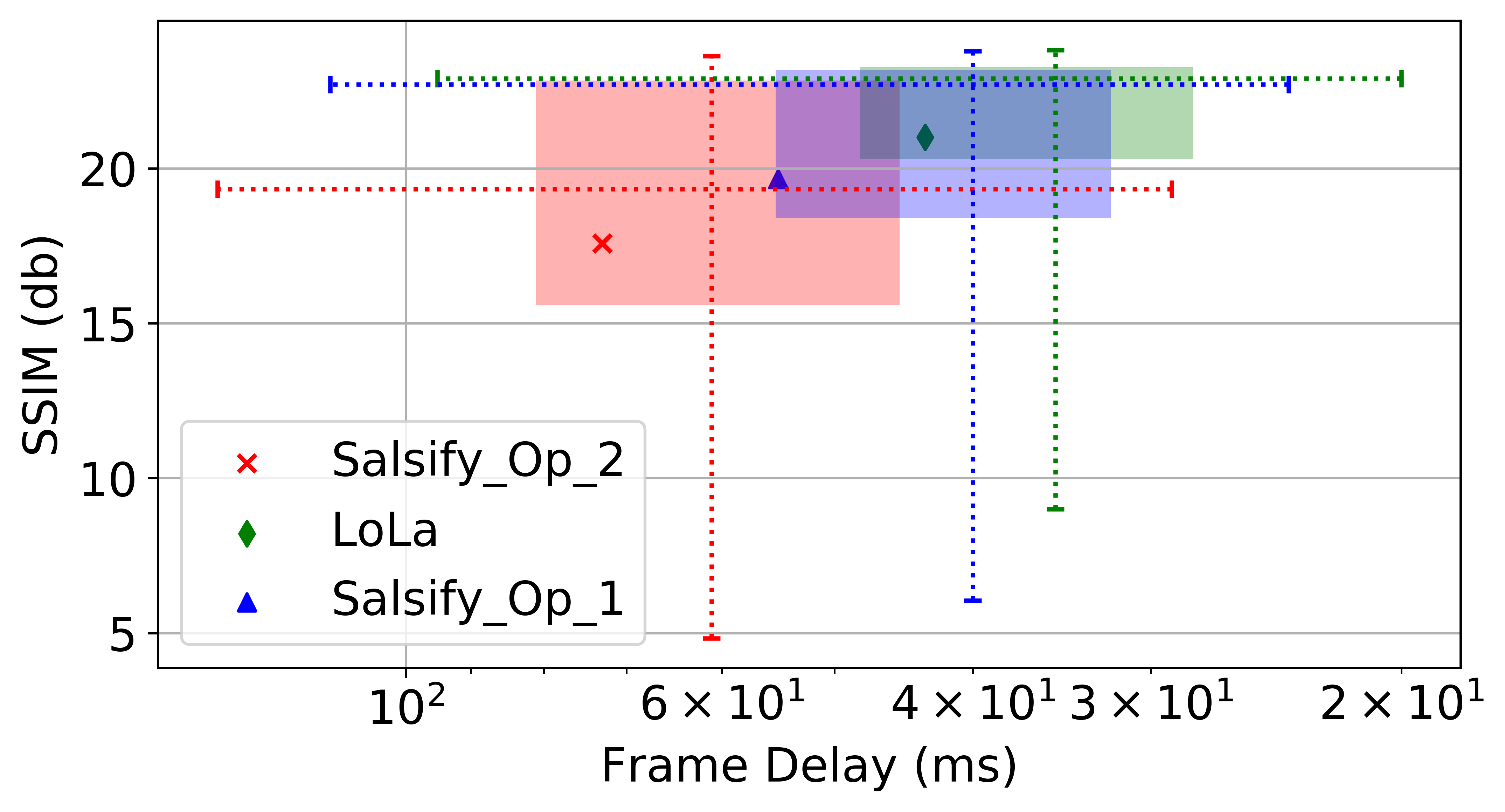}
    \end{subfigure}
    \begin{subfigure}[b]{0.33\textwidth}
        \centering
        \includegraphics[width=1\linewidth]{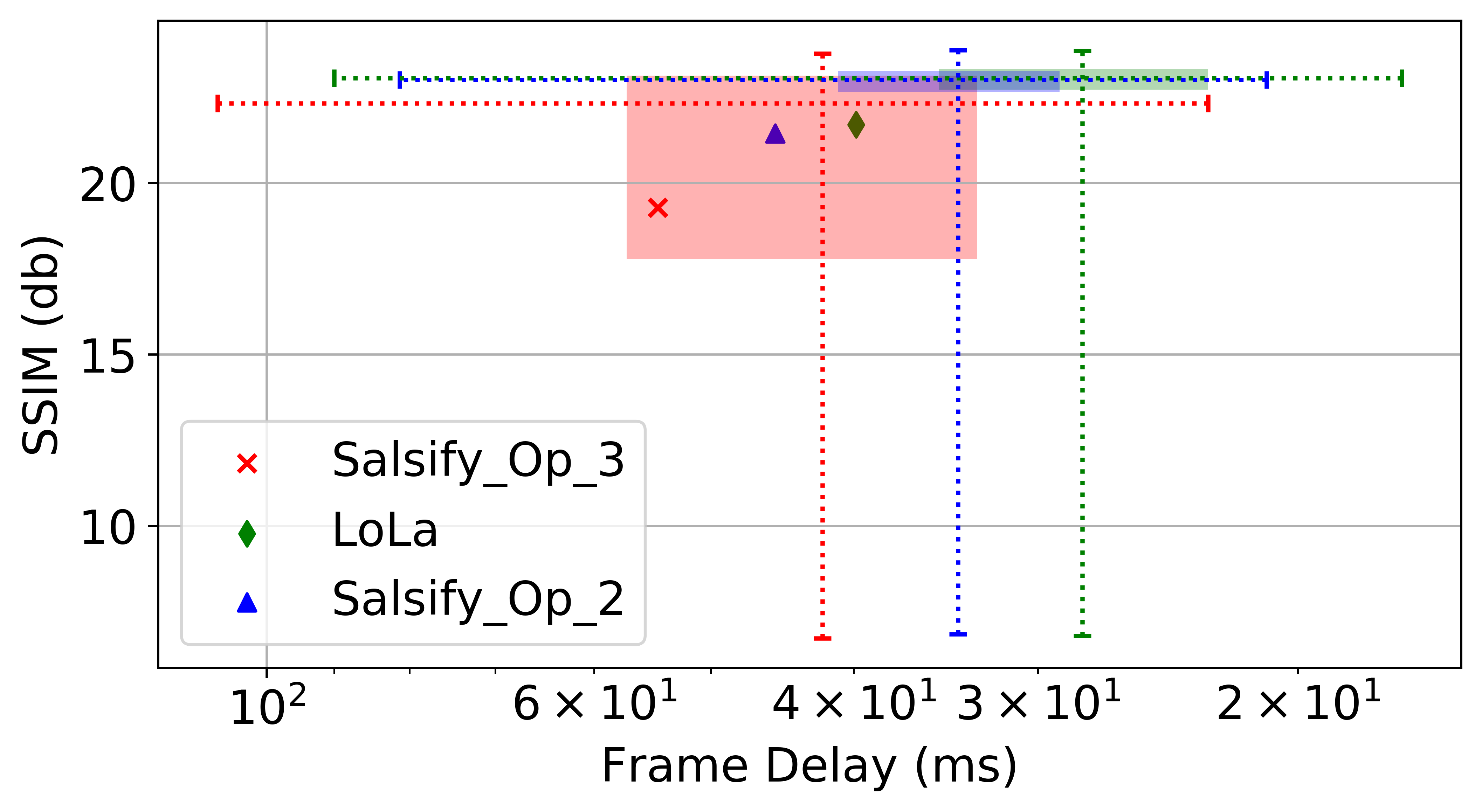}
    \end{subfigure}
    \caption{\sysname{} vs. Salsify in Outdoor Mobile Setting}\label{fig:mobile_results}
\end{figure*}

We replay the collected network traces using our CellNem replay tool three times,
once while running \sysname{} over the pair of network traces, and twice
with Salsify running over each one of the traces. This allows us to compare the
performance of \sysname{} against Salsify running on either one of the available
network traces. Our choice of conducting the comparative analysis against
Salsify stems from the fact that Salsify is the best performing single-trace
video conferencing solution outperforming WebRTC, Skype, Hangout, and Facetime.
 
During the evaluation, we pass video frames from a 1-minute video trace
captured using Logitech Brio 4K Pro Webcam recorded at 720p resolution and
60 frames-per-second. All of our experiments
are conducted using an ubuntu 18.04 machine with 12 CPUs, 2.6Ghz, and 30GB RAM.
We report the average, 5th, 25th, 75th, and 95th percentile of the per-frame Structural
Similarity Index (SSIM) in decibels (db) and the end-to-end frame-delay achieved in
each run. The end-to-end frame delay is measured as the time 
elapsed between when a frame is captured and passed to the encoder at the
sender end until the time the frame is decoded and displayed at the receiver
end. Our results over a sample of the stationary and mobile cellular network
traces are illustrated in \figref{stationary_results} and
\figref{mobile_results}, respectively.

Overall, we observe that \sysname{} outperforms the state-of-the-art single
trace video conferencing solution while providing a 2 to 5 db points
improvement on average in the SSIM in the stationary setting and 0.3 to 4 db
improvement in the mobile setting, all while keeping the 95th percentile
of the frame delay near or below the 100 milliseconds threshold. This
demonstrates the limitation of single trace solutions to handle cellular link
outages and fluctuations, and attest to \sysname{}'s ability to efficiently
utilize multiple cellular interfaces to provide an improved video conferencing
experience.

\balance\section{Conclusion}
\label{sec:conclusion}

In this manuscript, we presented \sysname{}, the first system designed
to support video conferencing over multiple cellular interfaces.
\sysname{} is integrated with the state-of-the-art video codec to
enable per-frame adaptation of frame resolution. Moreover, \sysname{} is
equipped with a quick feedback loop which allows to quickly and accurately
balance the load across the available cellular interfaces. Using real
network traces collected in stationary and outdoor environments, we were
able to demonstrate the need and the merit of \sysname{} in providing
higher perceived quality for video conferencing sessions over state-of-the-art
single trace solutions. For future work, we aim to explore duplication and
forward error correction strategies to further optimize the use of multiple
cellular interfaces.

\begin{acks}
This material is based upon work supported by the National Science Foundation under Grant No. CNS-2223556.
\end{acks}

\newpage
\bibliographystyle{ACM-Reference-Format}
\balance\bibliography{paper}

\end{document}